\documentclass[rnote]{aa} 
\usepackage{graphicx}
\usepackage{txfonts}
\def\Sum{N_{\rm tot}}
\begin{document}
   \title{Lanthanides and other spectral oddities in 
\mbox{a Centauri}\thanks{Based on observations collected at the
European Southern Observatory, Paranal, Chile (ESO programmes
65.L-0316(A), 073.D-0504(A), and 076.B-0055(A))}}

   \subtitle{Ce {\sc iii}, Nd {\sc iii}, Kr {\sc ii}, 
and broad absorption features}

  \author{C. R. Cowley
          \inst{1}
          \and
          S. Hubrig\inst{2}
          \and
          J. F. Gonz\'{a}lez\inst{3}
          }

   \offprints{C. R. Cowley}

   \institute{Department of Astronomy, University of Michigan,
              Ann Arbor, MI 48109-1090, USA\\
              \email{cowley@umich.edu}
         \and
             AIP, An der Sternwarte 16, 14482 Potsdam, Germany\\
              \email{shubrig@aip.de}
          \and
             Instituto de Ciencias Astron\'{o}micas, del la Tierra y del Espacio,
             Casilla 49, 5400 San Juan, Argentina  \\
             \email{fgonzalez@icate-conicet.gob.ar}}

   \date{Received month date, 2010/  Accepted month date year}

 
  \abstract
  %
   {There is considerable interest in the helium variable a Cen
   as a bridge between helium-weak and helium-strong CP stars.} 
   {We investigate {Ce {\sc iii}} and other possible lanthanides in
    the spectrum the of hottest chemically peculiar (CP)
    star in which these elements have been found.
    A {Kr {\sc ii}} line appears within a broad absorption which
    we suggest may be due to a high-level transition in 
    \mbox{C {\sc ii}}.}
   {Wavelengths and equivalent widths are measured on high-resolution
    UVES spectra, analyzed, and their phase-variations investigated.}
   {New, robust identifications of {Ce {\sc iii}} and {Kr {\sc ii}} 
    are demonstrated.  
    {Nd {\sc iii}} is likely present.  A broad absorption 
    near $\lambda$4619 is present at all 
    phases of a Cen, and in some other early B stars.}
   {The presence of lanthanides in a Cen strengthens the view that
    this star is a significant link between the cooler CP stars
    and the hotter helium-peculiar stars.  Broad absorptions in a Cen 
    are not well explained.}

   \keywords{Stars: chemically peculiar --
                stars: variable --
                stars: individual(a Cen) --
                line: identification}

   \maketitle

\section{Introduction\label{sec:intro}}
Upper main sequence stars exhibit a rich variety of chemical
peculiarities.  The anomalies seem significantly constrained
by the temperature domains within which the given peculiarities
are found.  For example, stars with Am characteristics have 
temperatures below 10000K, while the Hg-Mn stars are found at
higher temperatures.  It is well known that the ability to find
such peculiar objects spectroscopically is greatly enhanced 
for objects within their known and characteristic temperature
domains.  Increasing interest attaches to instances where
the abundance or physical characteristics of one peculiar species
is found across the boundary to another (Wahlgren 2004).

The helium variable, a Cen (HR 5378, HD 125823) is an interesting
example of an object that crosses traditional boundaries.  It
has an effective
temperature in the range 19,000 to 20,000K (Bohlender, Rice,
and Hechler 2010, henceforth BRH).  This 
would place it among the helium-strong chemically peculiar
stars.  Yet the star has long been known to vary in type from
helium strong to helium weak (cf. Gray and Corbally 2009).
BRH show that this variation occurs because of helium-rich 
and helium-poor areas on the stellar surface, which they have
mapped using Doppler imaging.  Interestingly, BRH identify
$^3$He, found in other helium-weak stars such as the highly
unusual phosphorus-gallium star 3 Cen A (Castelli, et al. 1997,
Adelman \& Pintado 2000).  

In the present note, we announce the identification of the
lanthanide rare earth spectrum Ce {\sc iii} along with the possible 
presence of Nd {\sc iii}.  These results would associate a Cen with the 
cooler, traditional magnetic CP stars.  
We are unaware of the identification of any lanthanide spectrum
in a star with this high a temperature.  

\section{Observations\label{sec:obs}}

Several UVESPOP spectra (Bagnulo, et al. 2003) of a Cen were measured 
for wavelengths.  The spectra were all obtained on 3 March 2001, at
phase 0.089 according to the ephemeris of Catalano \& Leone (1996).
This is near the maximum of the helium line strengths.  Most of
the present results rest on these spectra.  Additional UVES spectra
were examined to investigate phase variations.  These were all
obtained from 3 to 13 May of 2005, and span the 8.8-day  period of
the star.

\section{Background\label{sec:back}}

Norris (1971) reported the identification of 
C {\sc ii}, N {\sc ii}, O {\sc ii},
Ne {\sc i}, Mg {\sc ii}, Al {\sc iii}, Si {\sc ii}, Si {\sc iii}, 
S {\sc ii}, and Fe {\sc iii}  in the spectrum of \mbox{a Cen}.
Additionally, Fe {\sc ii} and Sr {\sc ii} were judged to be weakly present.
We can confirm these findings.  Hubrig and Gonzalez (2007) identified
additional spectra: Cl {\sc ii}, Ca {\sc ii}, Mn {\sc ii}, 
and Ar {\sc ii}.  They
found Mn {\sc ii} lines near 6122 \AA\, to be in emission from phases
0.260 to 0.784.  At phase 0.089, when the UVESPOP spectra were
obtained, the Mn {\sc ii} absorption spectrum is very weak. 
However, we could
find several of the very strongest lines in Mn {\sc ii} Multiplet 3.

We also find numerous weak Ti {\sc ii} lines.  Ti {\sc iii} may be present
at threshold, but cannot be confirmed.  We find an ``average''
(see \S\ref{sec:anal})
$\log({\rm Ti}/\Sum) = -6.3 \pm 0.13$ standard deviation (sd).  
This is an excess of 0.8 dex
above solar.

The most unusual identification by Hubrig and Gonzalez is that of
Cl {\sc ii}, which we confirm.  The strong  presence of
this spectrum from ground-based spectra is not unprecedented
(Cohen, et al. 1969, Sadakane 1992), but not common.  Oddly, 
Cl {\sc ii} is weak or absent in 3 Cen A and HD 65949, stars with
strong P {\sc ii}, but strong in HR 6870, which has modest
P {\sc ii} (Cowley, et al. 2010, Little 1974, 
Collado \& L\'{o}pez-Garcia 2009).
The capricious behavior of two neighboring odd-Z elements
(P: Z=15, Cl: Z=17) in similar stars
argues strongly for a chemical explanation.

\section{Analysis\label{sec:anal}}

 We measured 2492 wavelengths from the UVESPOP spectrum 
of \mbox{a Cen}.  These
were analyzed by {wavelength} coincidence statistics 
(WCS, Cowley \& Hensberge 1981).  WCS gives a Monte Carlo
estimate of the probability that the wavelength coincidences
are due to chance.  This probability, also called the ``significance''
of the result, depends on the tolerance
for a coincidence, $\Delta\lambda$.  In the present case, 
we used $\Delta\lambda = 0.06$\AA.  
In addition to elements found in older studies, the WCS
showed a highly significant result ($<$0.0004) for Ce {\sc iii}, along with 
highly suggestive results for Nd {\sc iii} and Kr {\sc ii}.  

BRH find abundance variations in helium,
nitrogen, oxygen, and iron of 2 to 3 dex in different locations
on the surface of a Cen. In view of these results, any abundance
based on an LTE calculation with a homogeneous model atmosphere 
has only semi-quantitative meaning.  
Nevertheless, we made formal calculations,
based on a plane-parallel model atmosphere with 
$T_{\rm eff} = 19500$K, and 
$\log(g) = 4.25$ for several elements.  
The relevant LTE codes have been in use by CRC and colleagues at
Michigan for several decades (see Cowley, et al. 2000, \S 7,
and references therein).
The {abundances must be considered}
a kind of average for the photosphere, with the nature
of the average being unspecified.  When the surface  has
been mapped, as in BRH, the nature of the average may 
become clarified.  We refer here to ``average'' abundances
with this qualification, and appropriate caveats for the
use of LTE.

\section{Ce {\sc iii} \label{sec:ce3}}

Table~\ref{tab:ce3} shows laboratory ($\lambda$) and stellar wavelengths
($\lambda^*$), excitation potentials,
equivalent widths, and LTE ``average'' abundance
calculations for 12  Ce {\sc iii} lines.   Oscillator strengths are
from the Li, et al. (2000) when available, or the 
DREAM site (Bi\'{e}mont, et al 2002).
\begin{table}
\caption{Ce {\sc iii} lines. \label{tab:ce3}}
\begin{tabular}{c c c r c}   \hline
$\lambda$&$\lambda^*$&$\chi$(eV)&$W$ [m\AA]&
$\log({\rm Ce}/\Sum)$ \\ \hline
3085.10& .07 &2.38&  8.2&    -8.31 \\
3121.56& .57 &2.38& 12.1&    -8.25  \\
3141.28& .29 &2.41&  5.5&    -8.36  \\
3143.97& .97 &2.41& 10.2&    -8.10  \\
3353.29& .32 &2.66&  9.9&    -8.21  \\
3427.36& .39 &2.38&  9.5&    -7.94  \\
3443.63& .66 &2.38& 10.9&    -8.02  \\
3454.39& .39 &2.40& 10.3&    -8.03  \\
3459.39& .41 &2.66&  7.7&    -8.28  \\
3470.92& .93 &2.41& 11.8&    -8.04  \\
3504.63& .67 &2.71& 11.2&    -7.85  \\
3544.06& .08 &2.71&  6.6&    -8.17  \\ \hline
\end{tabular}
\end{table}

The result, based on the lines in Table~\ref{tab:ce3}
is $\log({\rm Ce}/\Sum)=-8.13\pm 0.16$\,sd which
corresponds to an excess of cerium over solar of 2.3 dex
(Asplund, et al. 2009).
This result falls well within the range of cerium abundances
found for magnetic Ap stars (Ryabchikova, et al. 2004), and
for some HgMn stars (Adelman, et al. 2001ab).

\begin{figure}[ht]
\resizebox{\hsize}{!}{\includegraphics[angle=-00]{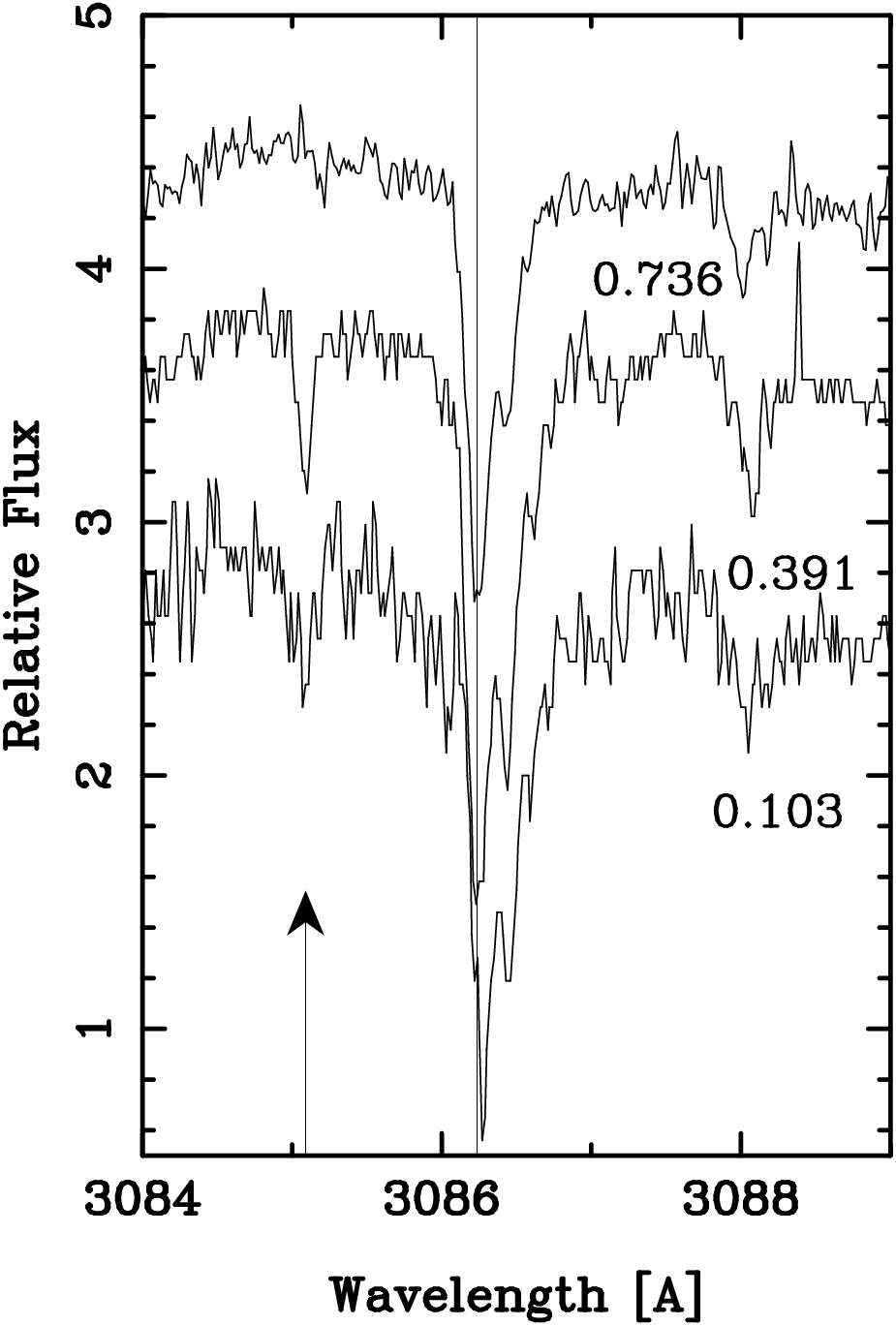}}
 \caption{Variations in a Cen of Ce {\sc iii} in $\lambda$3085.10
at three indicated phases.  The rest
wavelength is indicated by the arrow.  Spectra have been displaced
vertically for display purposes.  The length of the arrow, from
base to tip is 0.12 of the continuum.  The strongest feature is
a blend of Si {\sc iii}, $\lambda\lambda$3086.24, 3086.44, and 3086.67.
The absorption feature near $\lambda$3088 is Fe {\sc ii} 3087.97.  
\label{fig:5492fig1}}
\end{figure}

The Ce {\sc iii} lines
vary in strength and slightly in wavelength with phase, as is 
illustrated in Fig.~\ref{fig:5492fig1}.  

\section{Nd {\sc iii}, Kr {\sc ii}\label{sec:krnd}}
\subsection{Neodymium\label{sec:nd}}
The Nd {\sc iii} identification is marginal.  Of the six strongest lines
predicted with the DREAM oscillator strengths for $T = 19500$K,
5 are within 0.1\AA.  There is no indication at all of the missing
line, which probably means the others are present only 
{\it because} of blending--the blend strengthens the feature.
The blending would account for the wavelength distortions.
The strongest WCS result is 4 out of 16 strong
lines within 0.06\AA\, from a list provided to CRC by H. M. Crosswhite 
privately in 1976.  The coincidences have a significance of 0.003.
These results are typical of a spectrum plausibly present
near the threshold of detectability, and requiring confirmation.

\subsection{Krypton\label{sec:kr}}

The Kr {\sc ii} identification is robust.
There are seven ``persistent'' (P) lines in the NIST Handbook 
(Sansonetti \& Martin 2003) within the wavelengths measured.
The intensities range from 1000 to 150.  The five strongest of these 
lines were measured, though
shifted in wavelength on the UVESPOP spectrum (phase = 0.89) 
by an average of 
0.06\,\AA, or about 4.5 km s$^{-1}$.  
The lines are shown in Table~\ref{tab:2}.  

\begin{table}[h]
\caption{Persistent Kr II lines.
\label{tab:2}}
\begin{tabular}{c c r c c c} \hline
$\lambda$&$\Delta\lambda$&Int&$\chi$(eV)&$W_\lambda$[m\AA]&
$\log({\rm Kr}/\Sum)$ \\ \hline
4355.48&.05&  100P&13.99& 6.8 &$-$6.01 \\
4619.17&.06&  300P&14.69& blend& -- \\
4658.88&.05&  700P&13.99& 3.0 &$-$6.18\\
4739.00&.07& 1000P&13.99& 5.7 &$-$5.98 \\
4765.74&.06&  300P&14.27& 4.5 &$-$6.13\\
4832.08&.07&  250P&14.27& 2.8 &$-$5.79 \\ \hline
\end{tabular}
\end{table}

Table ~\ref{tab:2} gives gives wavelengths and intensities (Int)
from the NIST 
Handbook of Sansonetti and Martin (2003), along with shifts
$\Delta\lambda$ from the laboratory positions.  Lower excitation
potentials, measured equivalent widths, and abundance estimates
are also given.
The five abundances are consistent
within the accuracy of determinations of the present kind:
$\log({\rm Kr}/\Sum = -6.02\pm 0.16$\,sd.   This is an
excess of 2.8 dex over the solar abundance, similar to the
Kr excess (2.9 dex) of the mid-late B star, HD 65949
(Cowley, et al. 2010).

An additional 5 non-persistent Kr {\sc ii} lines with intensities
from 100 to 250 were measured, or may be found on the UVES
spectra: $\lambda\lambda$4431.685, 4436.812, 4577.209, 4615.29,
and 4762.453.  Only $\lambda$4431 deviates from its laboratory
wavelength by more than 0.03~\AA, if we use a rest radial
velocity frame defined by the lines of Table 2.
Moreover, this deviation may 
be attributed, from the broad profile, to a blend.  The Kr {\sc ii}
identification is secure.

\begin{figure}
\resizebox{\hsize}{!}{\includegraphics[angle=-90]{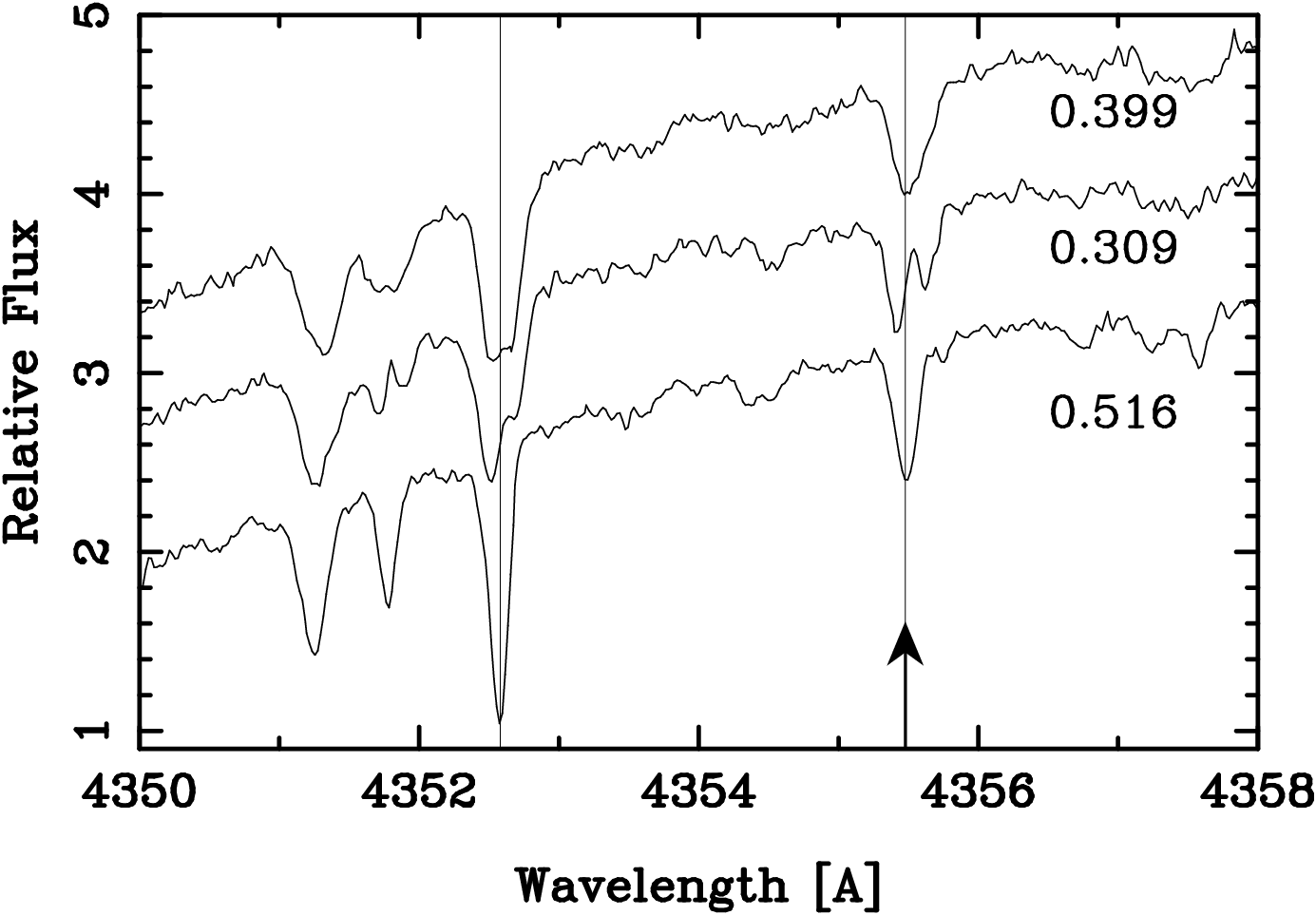}}
\caption{Kr {\sc ii} 4355.48 is indicated by the arrow.  Spectra 
for three indicated phases
have been displaced vertically for display purposes.  
We interpret the doubling of the Kr {\sc ii} at phase 0.309 as due to
approaching and receeding spots.  The length 
of the arrow, from
base to tip is 0.04 of the continuum.   The plot is in the wing
of H$\gamma$.
A vertical
line is drawn through Fe {\sc iii} $\lambda$4352.58.  
The O {\sc ii} and
Fe {\sc ii} lines $\lambda\lambda$4351.23 and 4351.77 appear  
shortward of this feature.
\label{fig:5492fig2}}
\end{figure}

Kr {\sc ii}, $\lambda$4355 is shown at three phases in
Fig.~\ref{fig:5492fig2}.  The splitting at phase 0.309 (111$^\circ$) 
as well as the shifts of Table~\ref{tab:2} are reasonably 
interpreted as due to abundance patches combined with stellar
rotation.
BRH's Fig. 6, shows iron spots at phases 0.250
(90$^\circ$) and 0.375 (135$^\circ$) on leading and following 
hemispheres.  Note that
both the Kr {\sc ii} and the Fe {\sc ii} 
double at our phase 0.309.  The value
$v\cdot\sin(i) \approx 15$ km s$^{-1}$ is adequate to produce the
0.20\AA\,
splitting seen at phase 0.309.  
This interpretation assumes the krypton and iron spots are similarly
placed on the stellar disk.

\section{Broad absorption features\label{sec:broad}}
\subsection{The broad absorption near $\lambda$4619\label{sec:broad46}}
\begin{figure}
\resizebox{\hsize}{!}{\includegraphics[angle=-90]{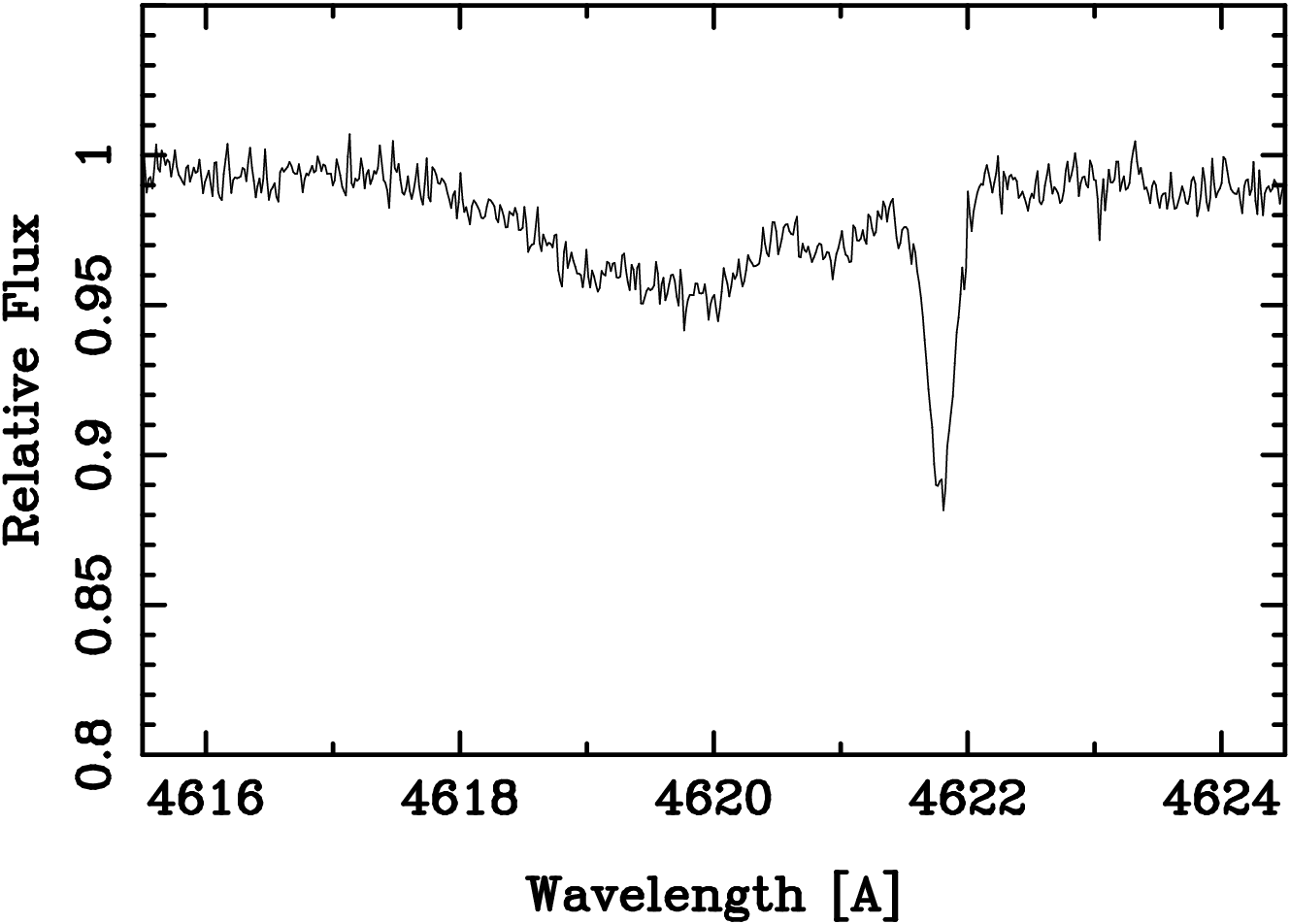}}
 \caption{Broad absorption near 
$\lambda$4619 in $\gamma$ Peg.  The UVES
spectrum, from the ESO archives was obtained on 24 November 2005.
The strong, narrow absorption is primarily N II $\lambda$4621.40.
\label{fig:5492fig3}}
\end{figure}

One of the persistent Kr {\sc ii} lines falls within a broad, shallow
absorption region.  The feature is present at all phases of a Cen,
and is seen in some other early B spectra.  
For example, it may be noted by
displaying the top 5 or 10\% of the continuum of UVESPOP spectra
of HD 133518 (B2 IV) and HD 89587 (B3 III).   In Fig.~\ref{fig:5492fig3}
we illustrate the broad absorption in this region of the spectrum 
of the well-known B2 IV star, $\gamma$ Peg.

\begin{figure}
\resizebox{\hsize}{!}{\includegraphics[angle=-90]{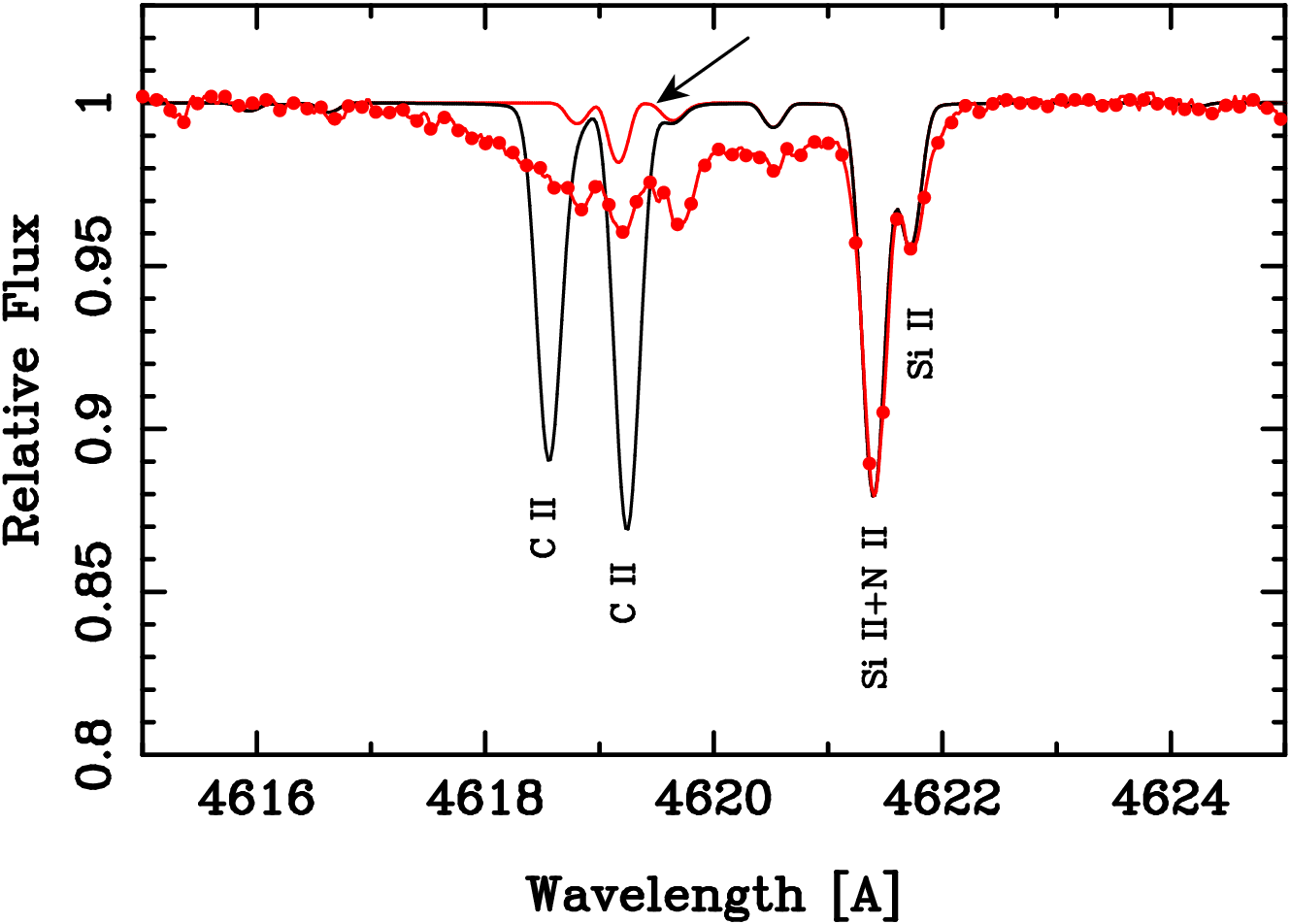}}
 \caption{The broad absorption near $\lambda$4619.  Observations
are in gray (red online) with dots.  Two sets of calculations are
shown.  The solid black line shows calculations including carbon,
but with $no$ krypton.
The fainter gray line (arrow, red online) is a calculation with no carbon, 
but including krypton with an abundance determined from five
other Kr {\sc ii} lines.
\label{fig:5492fig4}}
\end{figure}

Fig.~\ref{fig:5492fig4} shows an attempt to synthesize the
region.  It is clear that the atomic lines used in the calculation
are unable to reproduce the feature.  One of the three lines within 
the broad region coincides with Ti {\sc iii}, $\lambda$4619.79, but to
fit the feature, we need an abundance that yields
a strong Ti {\sc iii} line at 4615.93, which 
does not appear at all in the stellar spectrum.  Neither line appears
if we use the average Ti abundance from \S\ref{sec:back}.

It is tempting to assign the broad feature to C {\sc ii}.  The two prominent
calculated C {\sc ii} lines in Fig.~\ref{fig:5492fig4} were made using a carbon
abundance of $\log({\rm C}/\Sum) = -3.4 \pm 0.11$ which was determined 
from 19 other
C {\sc ii} lines.  If the feature is not due to C {\sc ii}, 
why do these lines
not appear? \, C {\sc ii} $\lambda\lambda$4618.56 and 4619.25 arise from a 
high-level multiplet (24.79 eV)
2s2p($^3$P)3d$^2$F$^\circ$---2s2p($^3$P)4f\,$^2$G, with
lower levels above the first ionization energy of C {\sc ii}
(24.38 eV).  Moreover,
the total absorption of the broad feature, 50 to 60m\AA, is comparable
to that of the calculated C {\sc ii} lines.  
However, the centroid of the broad feature is shifted away
from the C {\sc ii} lines, and the relevant energy levels for the lines
are given to the hundredth of a cm$^{-1}$.  This does not suggest
levels broadened by autoionization.  The broad feature requires an
explanation that must realistically include an NLTE calculation.

\subsection{The putative broad Fe {\sc i} absorption\label{sec:underk}}

Underhill and Klinglesmith (1973, see also Underhill et al. 1975)
found broad absorption features in a Cen which they attributed to
Fe {\sc i}.  Norris and Baschek (1974) were unable to confirm 
the presence of Fe {\sc i} in their spectra.  These studies were based
on photographic material, but averaged to reduce the noise.  While
we  also cannot confirm the 
association with Fe {\sc i}, we do see broad 
features near two of the positions of strong Fe {\sc i} lines:
$\lambda\lambda$4045, and 4383.  These absorptions were
readily detectable on UVES spectra taken at phases 0.795 and 
0.944.  The case that these absorptions are due to Fe {\sc i},
does not seem strong, but the broad absorptions in a Cen
deserve further attention.

\section{Conclusions}
The presence of lines of lanthanide rare earths in a star as hot as
a Cen indicates that these anomalies can ``cross a boundary''
(Wahlgren 2004) into a  temperature domain
where they are not normally recognized.
The likelihood that this is because of the difficulty of observing
lines from these elements rather than their absence is important to
note.
\begin{acknowledgements}
CRC thanks colleagues at NIST,
M. Dimitrijevi\'{c} and M. Bautista for useful correspondence
concerning the broad absorption feature near $\lambda$4619.
G. M. Wahlgren kindly pointed out the reference for Ce {\sc iii}
oscillator strengths.
This research has made use of the SIMBAD database, operated at
CDS, Strasbourg, France.  Our calculations made extensive use 
of the VALD atomic data base (Kupka, et al. 1999).   Special
thanks are due to the ESO staff for the UVES public data archive.
 
\end{acknowledgements}

\end{document}